\documentclass[journal]{IEEEtran}
\ifCLASSINFOpdf
\usepackage[pdftex]{graphicx}
\else
\fi
\usepackage{siunitx}
\usepackage{algorithmic,algorithm}
\usepackage{physics}
\usepackage{color}

\begin{document}

\title{Area Optimisation of Two Stage Miller Compensated Op-Amp in 65\,nm using Hybrid PSO}

\author{Ria Rashid and Nandakumar Nambath
\thanks{Ria Rashid and Nandakumar Nambath are with the School of Electrical Sciences, Indian Institute of Technology Goa, Ponda - 403401, India (e-mail: ria183422005@iitgoa.ac.in, npnandakumar@iitgoa.ac.in).}}

\markboth{}%
{Shell \MakeLowercase{\textit{et al.}}: Bare Demo of IEEEtran.cls for IEEE Journals}

\maketitle

\begin{abstract}

  Analog circuit design can be formulated as a nonlinear constrained optimisation problem that can be solved using any suitable optimisation algorithms. Different optimisation techniques have been reported to reduce the design time of analog circuits. A hybrid particle swarm optimisation algorithm with linearly decreasing inertia weight for the optimisation of analog circuit design is proposed in this study. The proposed method is used to design a two-stage operational amplifier circuit with Miller compensation. The results show that the proposed optimisation method can substantially reduce the design time needed for analog circuits.
  
\end{abstract}


\IEEEpeerreviewmaketitle

\section{Introduction}

  \IEEEPARstart{A}{nalog} circuit design has become an integral part in interfacing the real world signals with digital signal processing applications. There may be multiple analog interfaces to process signals such as voice, touch, motion, image, video, etc., present in a single chip. However, because of the shrinking device dimensions and the reducing supply voltages etc., analog circuit design has become increasingly complex. The highly nonlinear relationship between the circuit performance and design parameters has made analog design heavily dependant on the designer's intuition and experience \cite{1,2}. To reduce this dependency, researchers are actively looking at automated analog circuit design as a solution. To automate a design, circuit sizing can be formulated as a nonlinear constrained optimisation problem that can be solved using optimisation algorithms.

  Many optimisation techniques have been reported in literature for the optimisation of analog circuits \cite{3}. Gradient-descent based optimisation techniques \cite{4} that have been reported for analog circuit optimisation involve the calculation of derivatives, and the optimal solution obtained is highly dependant on the initial guess for the design variables. Convex optimisation techniques reported for the automation of analog circuit design \cite{5} result in a  global optimum solution, but it requires a thorough understanding of the transistor models to design the constraints. Since the state-of-the-art transistors have really complex mathematical models, the problem becomes even more challenging. In recent studies \cite{6,7}, Bayesian optimisation methods have been reported for automated analog circuit sizing. Another class of algorithms, called the evolutionary algorithms (EAs) have been reported to be used in analog circuit optimisation. EAs can give global solutions for complex optimisation problems. Moreover, EAs do not require the calculation of derivatives and can be applied to optimise complex systems. One such popular heuristics-based EA is particle swarm optimisation (PSO). PSO has been implemented in a wide range of real world problems because of its advantages such as robustness, simple representation, lesser number of adjustable parameters, ease of implementing parallel computation, and short computational time \cite{8}.

  PSO has been reported to be used for the optimisation of circuit sizing problems in different studies. Studies also show that PSO can be combined with other optimisation techniques for area optimisation of analog circuits \cite{9,10}. In \cite{11}, a modified version of PSO called hierarchical PSO has been used to optimise the component sizing of various analog circuits and has been found to give solutions with better repeatability when compared to PSO. PSO algorithm has been reported in \cite{12} to be used for the area optimisation of a two-stage operational amplifier (op-amp). Simulation results have proven that the PSO-based design met all design specifications and also gave a better result in terms of area with respect to the convex optimisation method. A modified PSO algorithm has been used for area optimisation of a complementary metal oxide semiconductor (CMOS) differential amplifier while considering design specifications of gain, phase, power dissipation, area occupied, etc. in \cite{13}. A craziness based PSO reported in \cite{14} has been found to give better results for the area optimisation of a CMOS two-stage op-amp than other reported techniques.

  In this paper, a hybrid PSO with linearly decreasing inertia weight is proposed for the area optimisation of analog circuits. A particle generation function is used to generate the particles in the swarm together with a survivability test to ensure that the particles meet all the circuit requirements. In this study, the decision vector has been formulated with transistor widths and the bias current. The circuit specifications are considered as the constraints for the optimisation problem. We have used the proposed optimisation technique to design a two-stage op-amp with Miller compensation.

\section{Overview}

  Formulation of the analog circuit optimisation problem and an overview of the standard PSO are given in this section.

\subsection{Problem Formulation}

  Analog design deals with trade-offs between various circuit parameters such as noise, linearity, gain, supply voltage, voltage swing, speed, input/output impedance, and power dissipation \cite{1}. Such trade-offs create many challenges in the analog circuit design procedure that requires experience and intuition to get optimal circuits. It can be said that such trade-offs along with specific circuit requirements make the analog design procedure a multidimensional optimisation problem, that has a solution space rather than a single solution. All analog circuit designs can be considered as optimisation problems with the design parameters as the design variables and the circuit specifications as constraints. The expression for the fitness function depends on the circuit parameters that need to be optimized for a particular analog design.
  
  The primary objective of an optimisation problem is to determine a vector $\vb*{x}=[x_1, x_2, x_3, \cdot\cdot\cdot ,x_n]$, called the position vector, that minimizes or maximizes a fitness function, $f(\vb*{x})$. $\vb*{x}$ is an $n$-dimensional vector, where $n$ represents the number of decision variables, the values of which have to be determined in the optimisation problem. The value of the fitness function, $f(\vb*{x})$, gives a measure of how good a particular solution represented by $\vb*{x}$ is. The set of all possible solutions constitute the search space for an optimisation problem.
  
\begin{figure}[!t]
    \centering
    \includegraphics[width=0.4\textwidth]{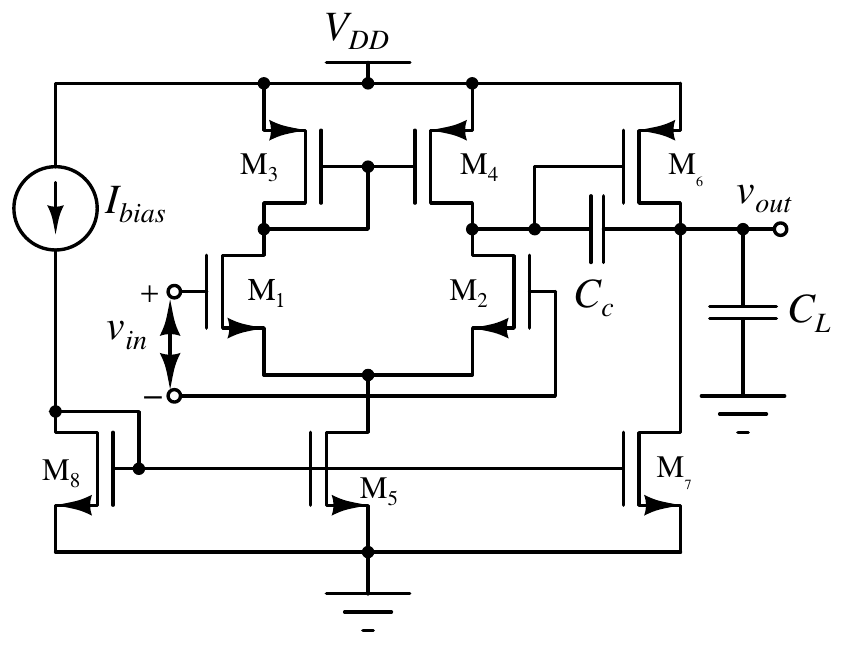}
    \caption{Schematic of a two stage op-amp with Miller compensation.}
    \label{fig1}
  \end{figure}

  The main goal of our work is to optimise analog circuits in terms of area i.e., optimal sizing of transistors. The proposed optimisation method is used to design a two stage Miller compensated op-amp. A schematic of the op-amp is shown in Fig.~\ref{fig1}. Its first stage comprises a differential amplifier, with an NMOS input stage consisting of M$_\text{1}$ and M$_\text{2}$ and a PMOS current mirror load consisting of M$_\text{3}$ and M$_\text{4}$. A current mirror consisting of M$_\text{5}$ and M$_\text{8}$ is used to bias the first stage. The second stage is a common source amplifier with a PMOS input stage, M$_\text{6}$ and an NMOS current source load, M$_{7}$.
  
  In Fig.~\ref{fig1}, we have assumed that the transistors M$_\text{1}$ and M$_\text{2}$, M$_\text{3}$ and M$_\text{4}$, and M$_\text{5}$ and M$_\text{8}$ are matched such that $W_1 = W_2$, $W_3 = W_4$, and $W_5 = W_8$. The bias current is denoted as $I_{bias}$. Since our objective is to minimize the chip area while meeting all other design specifications, the area of the circuit is chosen as the fitness function. The bias current and the widths of the transistors are taken as the decision variables for the optimisation problem. Specifications such as minimum voltage gain ($A_{v,min}$), minimum cut-off frequency ($f_{3dB,min}$), minimum unity gain-bandwidth ($UGB_{min}$), maximum power dissipation ($P_{max}$), minimum slew rate ($SR_{min}$), input common-mode range ($ICMR$), maximum area ($A_{max}$), minimum phase margin ($PM_{min}$), and maximum noise ($S_{n,max}$) are taken as constraints for this problem. The position vector, $\vb* x$, comprising the decision variables, and the fitness function, $f(\vb* x)$, can be expressed as
  \begin{equation}
    \label{eq1}
    \vb*{x} = [W_{1,2}, W_{3,4}, W_{5,8}, W_6, W_7, I_{bias}]~\text{and}
  \end{equation} 
  \begin{equation}
    \label{eq2}
    f(\vb*{x}) = \Sigma_{i=1}^{M}W_i\times L_i,
  \end{equation}
  respectively, where $M$ is the total number of transistors in the circuit and $W_i$ and $L_i$ are the width and length, respectively, of the $i^\text{th}$ transistor. For the circuit under consideration, $M = 8$.

\subsection{Overview of PSO}

  PSO algorithm, introduced by Kennedy and Eberhart in 1995 \cite{15}, is inspired by the intelligent collective behavior of a swarm of animals such as a flock of birds or a school of fishes. Typically, these swarms work towards a common goal where each member continuously alters its trajectory based on individual experience as well as that of other members. 

  In the standard form of PSO, the system is initialized with a population/swarm of particles, with each particle representing a potential solution. Initial positions of the particles are assigned by random values within the bounds of the solution space as defined in the optimisation problem. Trajectory and position of each particle change with each iteration where they are updated based on the particle's current position , the best position in its history, and the global best position that any of the particles in the swarm has ever attained. Let the position vector, $\vb*x_k$, denote the position of a particle,  and the velocity-vector, $\vb*v_k$, represent the velocity of the particle, where $k$ is the index of the particle. Throughout the iterations, each particle remembers its individual best position and the global best position, i.e., the best solution of the swarm. If the number of particles in the population is $N$ and the dimension of the search space is $D$, then the position of the $i^\text{th}$ particle in the $j^\text{th}$ iteration is represented as
  \begin{equation}
    \label{eq3}\vb*x_{i,j} = [x_{1}, x_{2},x_{3}, \cdots, x_{D}]_{i,j}.
  \end{equation}
  
  For every iteration, the velocity and position of each particle are updated using the equations
  \begin{equation}
    \label{eq4}
    \vb*v_{i,j+1} = w\vb*v_{i,j} + c_1r_1(\vb*p_{i,j} - \vb*x_{i,j}) + c_2r_2(\vb*g_{j} - \vb*x_{i,j})~\text{and}
  \end{equation}
  \begin{equation}
    \label{eq5}\vb*x_{i,j+1} = \vb*x_{i,j} + \vb*v_{i,j+1}
  \end{equation}
  where $w$ is the inertia weight, $\vb*v_{i,j}$ is the velocity of the $i^\text{th}$ particle in the $j^\text{th}$ iteration, $\vb*p_{i,j}$ is the best position of the $i^\text{th}$ particle so far, $\vb*g_{j}$ is the global best position among all the particles, $r_1$ and $r_2$ are two random numbers uniformly distributed in the interval [0,1], and $c_1$ and $c_2$ are the constriction factors that are used to control and constrict the velocities. The iteration is repeated until all the particles have converged to the position representing the optimum solution.
  
\section{Proposed Optimisation Method}

  Details of the proposed optimisation technique and its application in the design of a two stage op-amp with Miller compensation are discussed in this section. Results of proposed method used for the area optimisation of a differential amplifier circuit is reported earlier in \cite{16}.

\subsection{Particle Generation Function} 

  \begin{figure}[!t]
    \centering
    \includegraphics[scale=0.4]{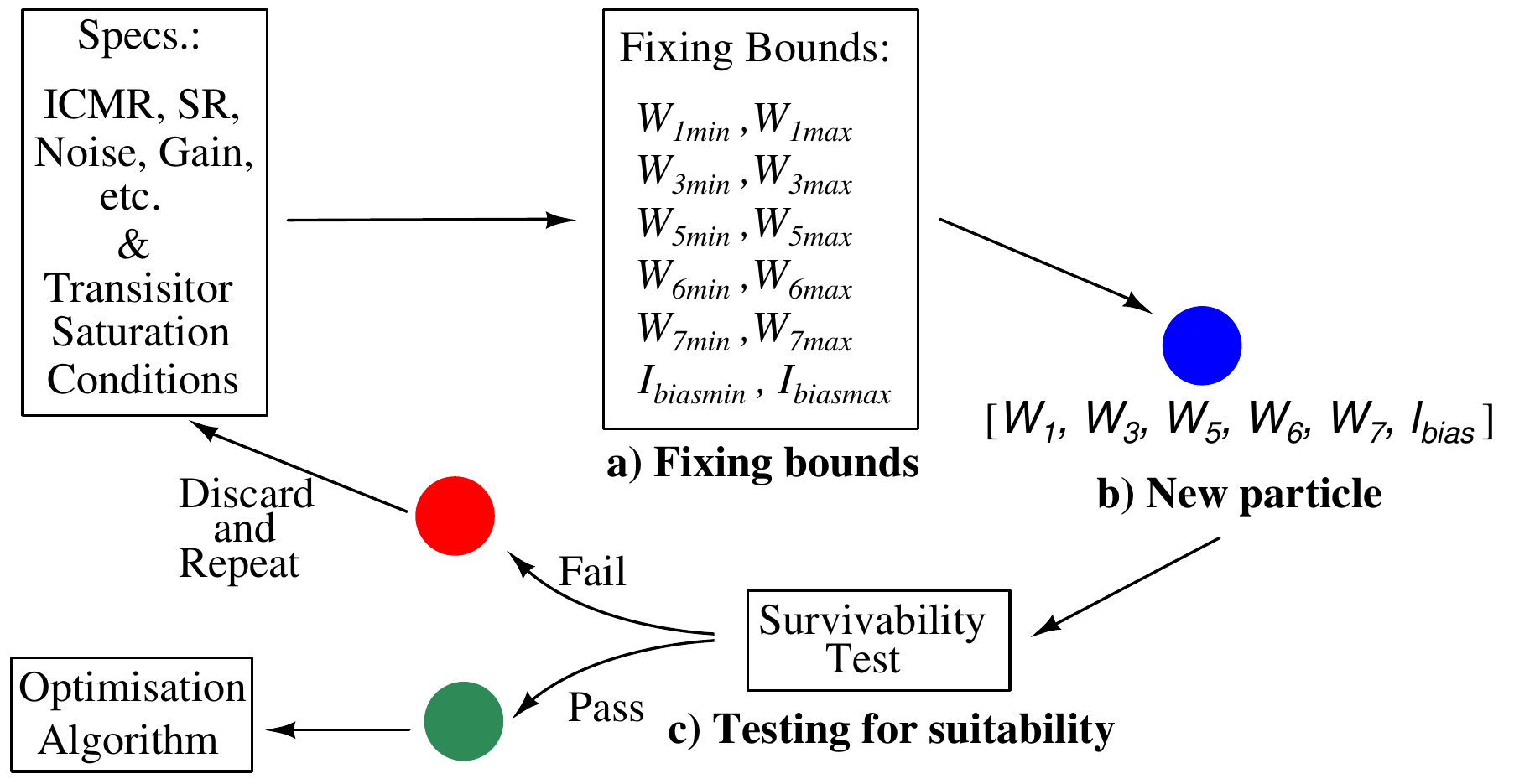}
    \caption{Steps involved in the particle generation function.}
    \label{fig2}
  \end{figure}
  
  Particles of the initial population as well as the subsequent iterations are generated using a particle generation function. A flow diagram depicting the particle generation function is given in Fig.~\ref{fig2}. As shown in step (a), the maximum and minimum bounds for the design variables are fixed using the design specifications listed in Table \ref{table1} and saturation conditions of the transistors using the I-V characteristics of the transistors. Random values are picked from these bounds for each of the design variables to create a particle, as illustrated in step (b). In step (c), the generated particle is made to undergo a survivability test. If the particle fails the test, it is discarded and the process is repeated until a suitable particle is obtained. Here, fixing the bounds for the design variables using the design specifications and the saturation conditions increases the probability of a particle passing the survivability test. This results in substantial reduction of time required to generate a particle.

\subsection{Survivability Test}

\begin{figure}[!t]
    \centering
    \includegraphics[scale=0.4]{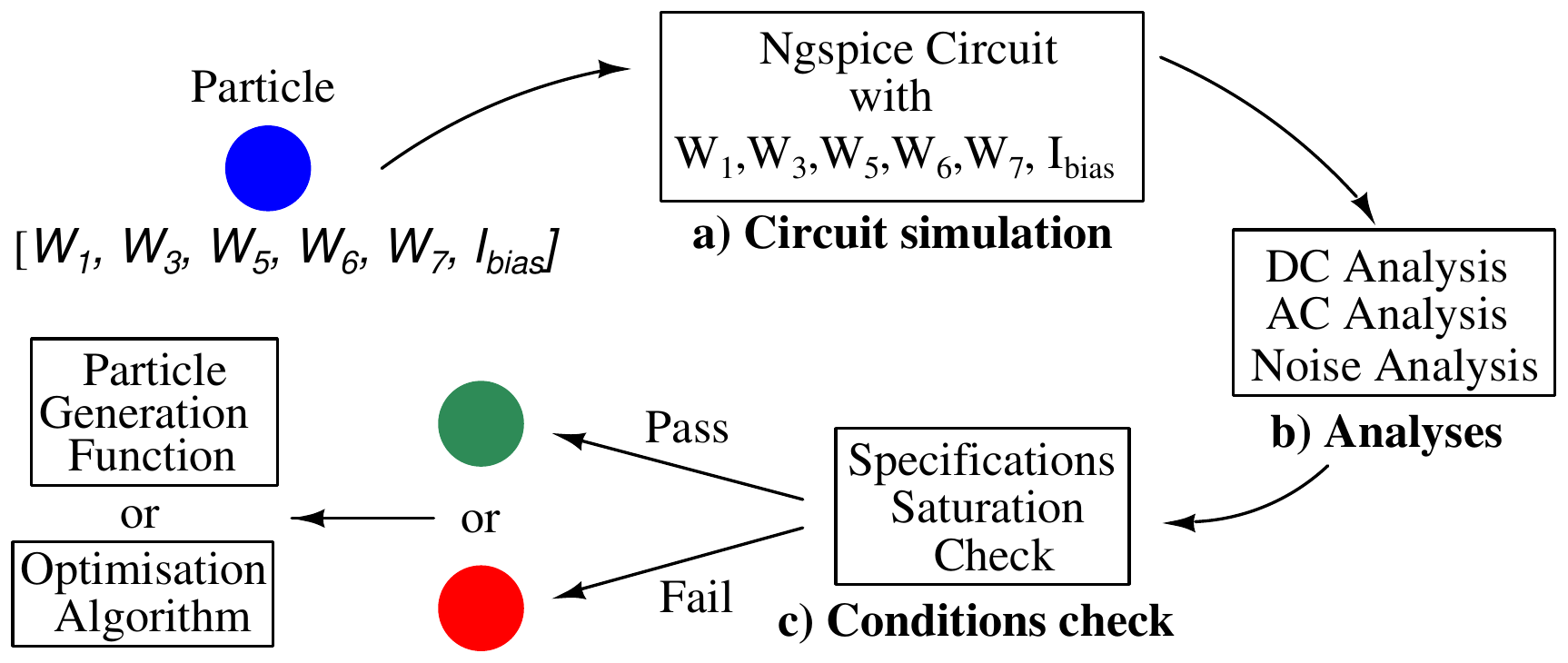}
    \caption{Steps involved in the survivability test.}
    \label{fig3}
  \end{figure}

  A particle is made to undergo a survivability test to determine whether it is meeting all the design specifications and saturation conditions of the transistors in the circuit. The survivability test is carried out using ngspice simulations. A flow diagram showing steps in the survivability test is given in Fig.~\ref{fig3}. The op-amp circuit that is being optimised is simulated using ngspice with the design parameters represented by each particle, as shown in step (a). In step (b), DC operating point analysis, AC analysis, and noise analysis are carried out. The circuit is then checked for saturation conditions as well as the required specifications, as illustrated in step (c).
 
\subsection{Hybrid PSO with Linearly Decreasing Inertia Weight} 

  Variants of the standard PSO have been used in multiple applications. One such variant is the  PSO with linearly decreasing inertia weight. In this variant, the inertia weight, $w$, is varied linearly between $w_{max}$ and $w_{min}$ over the iterations \cite{17,18}. This modification of the inertia weight ensures global exploration in the initial iterations, thereby maximising the probability of finding a global solution. A lower inertia weight towards the final iterations results in local exploration that helps in faster convergence to the optimal solution. This variant of PSO is found to give better results for our optimisation problem. Since the values of $w_{max}$ and $w_{min}$ are highly problem dependant, we have chosen the set of values giving consistent results for multiple runs with the best optimal value for the fitness function.
 
  \begin{figure}[!t]
    \centering
    \includegraphics[scale=0.24]{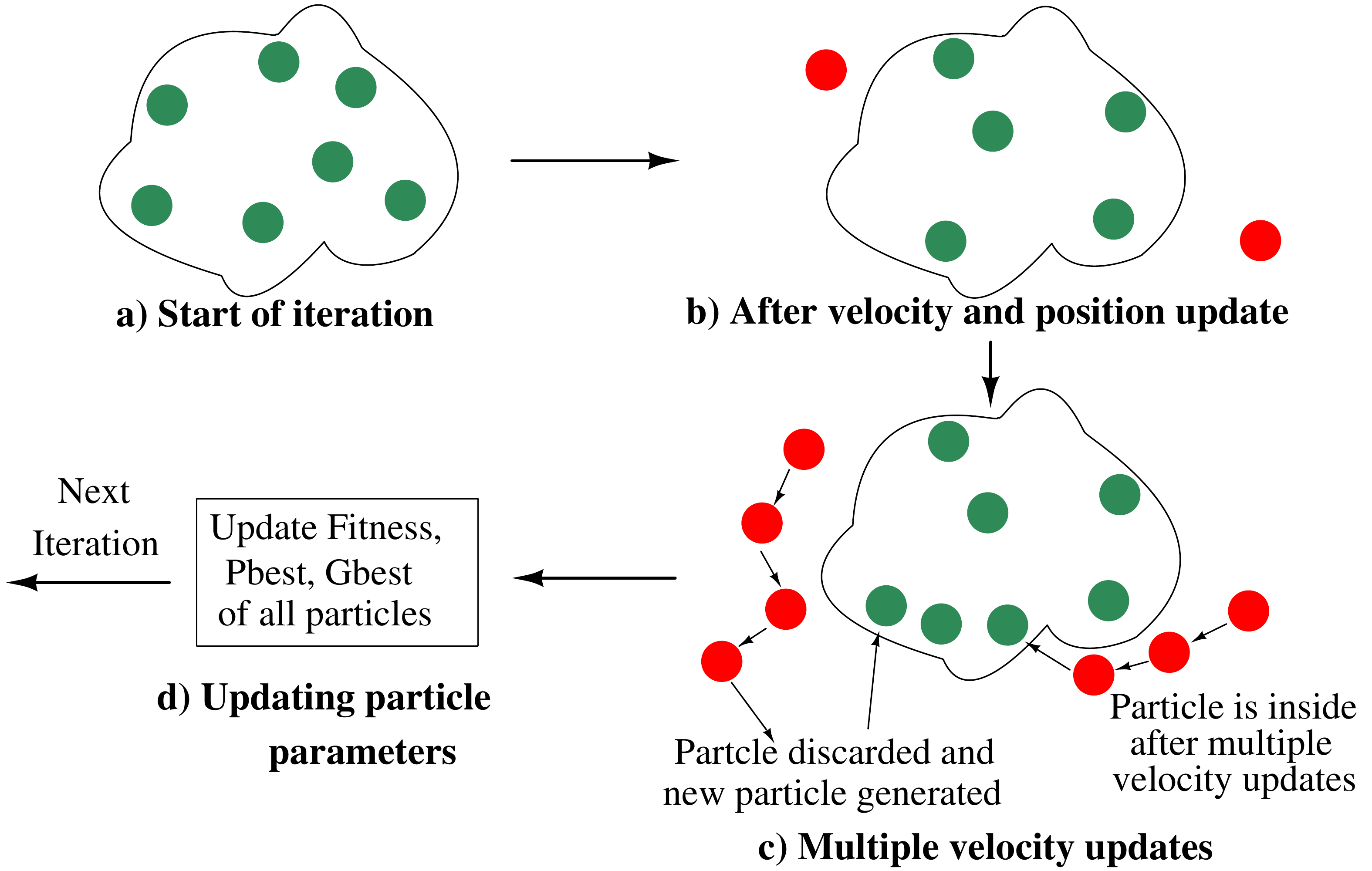}
    \caption{Pictorial representation of the proposed hybrid PSO.}
    \label{fig4}
  \end{figure} 
  
  A pictorial representation of the proposed hybrid PSO algorithm is given in Fig.~\ref{fig4}. The main difference of the proposed method from the standard PSO is how the velocity of each particle is getting updated in every iteration. For the proposed hybrid algorithm, the position vector and the fitness function are the same as (1) and (2). The position and velocity update equations are given in (4) and (5). The particles passing the survivability test are shown in green and inside the bounds, and those which are failing are shown in red and as out the of bounds in the figure. In this variant of PSO, the initial swarm is generated using the particle generation function, as shown in step (a) in the figure. In every iteration, after updating the position of each particle, the particle is made to undergo the survivability test. This is illustrated in step (b) in the figure, in which two particles are shown as out of bounds. Unlike standard PSO, for the particles out of bounds, velocity and position updates are carried out again using equations (4) and (5). These particles are subjected to the survivability test once more. If they fail the test, the above process is repeated either until suitable particles are obtained or the number of velocity updates reaches a predefined limit. Even after this limit, if the particles are unable to pass the survivability test, new particles generated using the particle generation function are introduced in their place. The multiple velocity updates for particles that are out of bounds is shown in step (c). This is carried out for all the particles throughout the iterations. Then the particle parameters are updated as shown in step (d). The next iterations are implemented in the same manner till the stopping criteria for the optimisation algorithm are met.
  
   \begin{algorithm}[t!]
    \caption{The proposed hybrid PSO.}
    \label{alg1}
    \begin{enumerate}	
	  \item \textbf{Particle generation function}
      \begin{enumerate}
        \item Generate particle.
        \item Conduct survivability test for the particle.
        \item If the test fails, go to 1a. Else, exit the function.
      \end{enumerate}
      \item \textbf{Main algorithm}
      \begin{enumerate}
        \item Initialise PSO parameters: $w_{max}$, $w_{min}$, $c_1$, $c_2$, $N$, $i=0$, $ite=0$, $count=0$, $maxite$, $maxcount$.
        \item Generate initial swarm.
        \begin{enumerate}
          \item If $i>N$, go to step 2c. Else, continue.
          \item Call particle generation function.
          \item Update $i=i+1$ and go to step 2b-i.
        \end{enumerate}
        \item Update initial velocity, fitness values, $pbest$, and $gbest$ of the swarm.
        \item Begin iterations: Set $ite=0$, $i=0$, $count=0$.
        \begin{enumerate}
          \item If $ite>maxite$, go to step 2e. Else, continue.
          \item Update inertia weight, $w$.
          \item Set $w=w_{min}+(w_{max}-w_{min}/maxite) \times ite$.
          \item If $i>N$, go to step 2d-x. Else, continue.
          \item Update velocity and position of the particle. Set $count=count+1$.
          \item If $count>maxcount$, call particle generation function.
          \item Conduct survivability test for the particle.
          \item If the test fails, go to step 2d-v. Else, continue.
          \item Update $i=i+1$ and go to 2d-iv.
          \item Update fitness value, $pbest$, and $gbest$
          \item Update $ite=ite+1$ and go to 2d-i.
        \end{enumerate}
        \item Update best fitness and the optimal solution.
      \end{enumerate}
    \end{enumerate}
  \end{algorithm}
  
  \begin{table}[!t]
    \begin{center}
    \caption{Specifications used for the two stage op-amp design.}
    \label{table1}
    \begin{tabular}[!t]{lc}
      \hline
      \hline
      \textbf{Specification} & \textbf{Value} \\
      \hline
      \hline
      Voltage gain ($A_{v}$) & $\ge$\SI{20}{\decibel} \\
      Power dissipation ($P$) & $\le$\SI{400}{\micro\watt} \\
      Slew Rate ($SR$) & $\ge$\SI{100}{\volt\per\micro\s}   \\
      Cut-off frequency ($f_{3dB}$) & $\ge$\SI{10}{\mega\hertz}   \\
      Unity gain bandwidth ($UGB$) & $\ge$\SI{100}{\mega\hertz}   \\
      Phase margin & $\ge$\SI{60}{\degree}   \\
      Input common-mode ($V_{in,CM}$) & 0.6\,V$\leq V_{in,CM}\leq$1.0\,V  \\
      Power spectral density ($S_n(f)$) & $\le$\SI{60}{\nano\volt\per\sqrt{Hz}} at \SI{1}{\mega\hertz}\\
      Area ($A$) & $\le$\SI{1}{\micro\meter^{2}}  \\
      Load capacitance ($C_L$) & \SI{200}{\femto\farad}  \\
      Transistor sizes ($W/L$) & 2 to 200  \\
      \hline
      \hline
    \end{tabular}
    \end{center}
  \end{table}
  
  \begin{figure}[!t]
    \centering
    \includegraphics[scale=0.182]{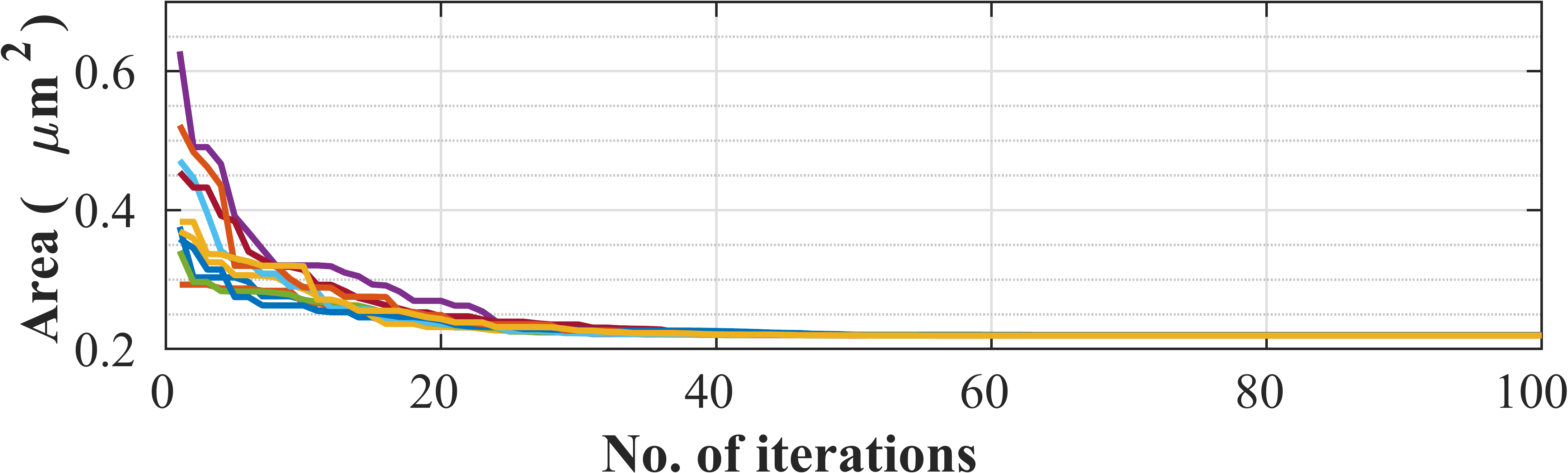}
    \caption{PSO convergence characteristics for 10 runs with 100 iterations.}
    \label{fig5}
  \end{figure}

  \begin{table}[!t]
    \begin{center}
    \caption{Optimum parameters obtained for two stage op-amp.}
    \label{table2}
      \begin{tabular}[!t]{ccccccc}
        \hline
        \hline
        $I_{bias}$ & $W_{1,2}$ & $W_{3,4}$&  $W_{5,8}$ & $W_6$& $W_7$\\
        \SI{29.7}{\micro\ampere}&\SI{266}{\nano\meter}&\SI{783}{\nano\meter}&\SI{126}{\nano\meter}&\SI{1115}{\nano\meter}&\SI{191}{\nano\meter}\\
        \hline
        \hline
      \end{tabular} 
    \end{center}
  \end{table}
   
  The repeated updates to the velocity in a single iteration in the proposed hybrid PSO is found to be helpful for this specific problem as the design space is highly complex, multidimensional, and nonlinear. After a particle has failed the survivability test, repeatedly updating the velocity and hence, its position, increases the probability of it moving to a suitable position in the search space. This also reduces the number of new particles introduced into the swarm, thereby reducing the time taken by the algorithm to converge. A pseudo-code of the proposed algorithm is given in Algorithm \ref{alg1}.
  
\section{Simulation Results}

  \begin{table}[!t]
    \caption{Optimum parameters obtained for the two stage op-amp.}
    \label{table3}
    \begin{center}
      \begin{tabular}{lccc}
        \hline
        \hline
        \textbf{Design Criteria}&{\bf Specifi-}&\multicolumn{2}{c}{\textbf{MC Results}} \\
        \cline{3-4}
        \textbf{}&{\bf cations}&\textbf{Mean}&\textbf{Std} \\
        \hline
        \hline
        $A_v$ (\si{\decibel})&$\geq20$&21.6&1.34\\
        $f_{3dB}$ (\si{\mega\hertz})&$\geq10$&13.9&1.6\\
        $UGB$ (\si{\mega\hertz})&$\geq100$&169.7&4.6\\
        Phase Margin $(^{\circ})$&$\geq60$&62.4&2.64 \\
        $SR$ (\si{\volt\per\micro\second})&$\geq100$&288&47  \\
        $P$ (\si{\micro\watt})&$\leq150$&89&20  \\
        $S_n(f)$@1MHz (\si{\nano\volt\per\sqrt{Hz}})&$\leq60$&53&0\\
        $S_n(f)$@10MHz (\si{\nano\volt\per\sqrt{Hz}})&-&19&0\\
        $CMRR$ (\si{\decibel})&-&35.5&9.1\\
        $PSRR+$ (\si{\decibel})&-&20.8&2.9\\
        $PSRR-$ (\si{\decibel})&-&49.2&9.4\\
        Settling time with 2\% tol. (\si{\nano\second})&-&5.4&0.84\\ 
        Settling time with 5\% tol. (\si{\nano\second})&-&4.3&0.49\\ 
        $\textbf A$ (\si{\micro\meter^{2}})&$\leq1$& \multicolumn{2}{c}{\textbf{0.22}} \\
        \hline
        \hline
      \end{tabular} 
    \end{center}
  \end{table}
 
  The design specifications used for the two stage op-amp are given in Table \ref{table1}. The circuit is designed in \SI{65}{\nano\meter} technology. The supply voltage, $V_{DD}$, is taken as \SI{1.1}{\volt}. The length of all the transistors is fixed to \SI{60}{\nano\meter}. The PSO parameters are taken as: $w_{min}=0.5$, $w_{max}=0.8$, and $c_1 = c_2 = 1.7$. To ensure a phase margin of 60$^\circ$, the value of $C_c$ is taken as 0.3 times $C_L$.
  
  The proposed optimisation algorithm is implemented in Matlab with the survivability test performed using ngspice simulations. The algorithm is run multiple times with swarm sizes of 10, 15, and 20 for 100 iterations in each run. From the simulations, it is observed that the variance of the converged value of the fitness function decreases with increasing swarm size. Hence, a swarm size of 20 is chosen. The PSO convergence characteristics for 10 runs with 100 iterations with a swarm size of 20 are shown in Fig. \ref{fig5}. From the runs, the best value in terms of the fitness function is taken as the final solution, and the circuit parameters corresponding to these values are found out. The design parameters of the final solution obtained are given in Table \ref{table2}. The optimal solution is found to meet all the required specifications in the ICMR of \SI{0.6}{\volt} to \SI{1}{\volt}. The results for the Monte Carlo simulations for the optimal design for 1000 runs across various PVT corners are summarized in Table \ref{table3}. 

  The comparison of results of the proposed method and other reported studies in 350\,nm \cite{12} and 180\,nm \cite{19} technologies for the two stage op-amp are summarised in Table~\ref{table4}. It can be seen that the proposed algorithm gives a better optimal area than the ones reported in \cite{12,19}. The proposed method has been used for the design of the op-amp reported in \cite{7} for an unconstrained optimisation in 180\,nm technology and the comparison with the reported Bayesian optimisation method is given in Table \ref{table6}. It shows that the proposed technique converges to a higher fitness function value showing a better performance. The proposed algorithm converges to the optimal value in less than 50 iterations for all the considered cases.
  
  \begin{table}[!t]
    \caption{Op-amp optimum parameters comparison with other work.}
    \begin{center}
      \label{table4}
      \centering
      \begin{tabular}[!t]{lcccccc}
        \hline
        \hline
        & \multicolumn{3}{c}{\bf 350\,nm Technology}&\multicolumn{3}{c}{\bf 180\,nm Technology}\\
         \bf Design & \bf Specs.&\bf PSO & \bf This &\bf Specs.&\bf PSO& \bf This\\
         \bf Criteria&\cite{12}&\cite{12}& \bf work &\cite{19}&\cite{19}&\bf work\\
        \hline
        \hline
        $A_v$ (\si{\decibel}) &$\geq60$& 63.8 & 66.43 &$>50$& 59.19&56.37\\
        $UGB$ (\si{\mega\hertz}) &$\geq3$& 5.53& 3.90 &$>10$&20.03 &11.70\\
        Phase Margin $(^{\circ})$&$\geq45$&66.55&45.0&$>45$&63.53&45.0 \\
        $SR$ (\si{\volt\per\micro\s})&$\geq10$& 11.3 & 11.0 &$>10$& 18.35 & 11.0\\
        $P$ (\si{\milli\watt}) &$\geq2.5$& 2.37 & 0.98 &$<1$&0.184 &0.040\\
        $CMRR$ (\si{\decibel}) &$>60$& 83.74 & 82.13 &$>60$& 67.08 & 62.95\\
        $PSRR+$ (\si{\decibel}) &$>70$&78.27 & 78.24 &$>60$&63.84& 60.04\\
        $PSRR-$ (\si{\decibel}) &$>70$&93.56 & 92.93 &$>60$&99.16&74.52\\
        $V_{in,CM,max}$ (\si{\volt}) &$\leq2$&1.75& 1.75 &-&-&-\\
        $V_{in,CM,min}$ (\si{\volt}) &$\geq-1.5$&-0.8&-0.8&-&-&-\\
        $V_{DD}$ (\si{\volt}) &-& 2.5 & 2.5 &-&1.8&1.8\\
        $V_{SS}$ (\si{\volt}) &-& -2.5&-2.5&-&0&0\\
        $C_L$ (\si{\pico\farad}) &-& 10& 10&-&-&1\\
        $\textbf A$ (\si{\micro\meter^{2}})&$\bf\leq300$& \textbf{265} & \textbf{134} &$\bf\leq300$&\textbf{28.52} & \textbf{1.45}\\
        \hline
        \hline
      \end{tabular} 
    \end{center}
  \end{table}
    
  \begin{table}[!t]
    \caption{Comparison with Bayesian Optimisation of two stage op-amp.}
    \begin{center}
      \label{table6}
      \centering
      \begin{tabular}[!t]{lcccc}
        \hline
        \hline
        \bf Algorithm & {\bf Best}&{\bf Worst}& {\bf Mean}& {\bf Std}\\
        \hline
        \hline
        MACE \cite{7}&690.36&690.27&690.34&0.03\\
        This work&2110.1&2100.1&2104.8&2.9\\
        \hline
        \hline
      \end{tabular} 
    \end{center}
  \end{table}

\section{Conclusion}

  In this paper, a hybrid PSO with linearly decreasing inertia weight suitable for the optimisation of analog circuit sizing is proposed. The proposed method is used to design a two-stage op-amp with Miller compensation in \SI{65}{\nano\meter} technology, and the simulation results are presented. The optimal solution obtained is found to meet all the required circuit specifications. The study shows that the proposed method can substantially reduce the design time required for analog circuits. The method can be used to optimise more complex analog circuits and for multi objective optimisation problems.
  
\bibliographystyle{IEEEtran}

\end{document}